\documentclass[12pt]{article}
\usepackage[margin=1in]{geometry}
\usepackage{color}
\usepackage{amsfonts}
\usepackage{amsmath}
\usepackage{natbib}
\usepackage{graphics}
\usepackage{graphicx}
\usepackage{rotating}
\usepackage{setspace}
\usepackage{enumitem}
\usepackage[font=small,labelfont=bf]{caption}
\numberwithin{equation}{section}

\begin{document}

\baselineskip = 18truept

\title{Wavelet-Based Scalar-on-Function \\ Finite Mixture Regression Models}
\author{Adam Ciarleglio and R. Todd Ogden \footnote{Adam Ciarleglio is a Postdoctoral Fellow, Department of Child and Adolescent Psychiatry, Division of Biostatistics, New York University, New York, NY (E-mail: Adam.Ciarleglio@nyumc.org).  R. Todd Ogden is Professor, Department of Biostatistics, Columbia University, New York, NY (E-mail: to166@columbia.edu).}}

\date{}
\maketitle
\doublespacing
\begin{abstract}
Classical finite mixture regression is useful for modeling the relationship between scalar predictors and scalar responses arising from subpopulations defined by the differing associations between those predictors and responses.  Here we extend the classical finite mixture regression model to incorporate functional predictors by taking a wavelet-based approach in which we represent both the functional predictors and the component-specific coefficient functions in terms of an appropriate wavelet basis.  In the wavelet representation of the model, the coefficients corresponding to the functional covariates become the predictors.  In this setting, we typically have many more predictors than observations.  Hence we use a lasso-type penalization to perform variable selection and estimation.  We also consider an adaptive version of our wavelet-based model.  We discuss the specification of the model, provide a fitting algorithm, and apply and evaluate our method using both simulations and a real data set from a study of the relationship between cognitive ability and diffusion tensor imaging measures in subjects with multiple sclerosis.

\end{abstract}

\doublespacing

\section{Introduction}
Regression models for scalar responses and functional predictors have become increasingly important tools for analyzing functional data.  There are many applications in which these models can provide an adequate description of the data at hand.  For instance, a classic example provided in \cite{fda2} uses a functional linear model to describe the association between total annual rainfall (scalar response of interest) and temperature over the course of the corresponding year (functional predictor).

Let $Y_i \in \mathbb{R}$ be the scalar response of interest for observation $i$, $i = 1,\ldots,n$ and let $X_i$ be a random predictor process that is square integrable  on a compact support $I \subset \mathbb{R}$ (i.e, $\int_I{X^2_i(t)}dt < \infty$).  The corresponding functional linear model (FLM) is given by:
\begin{equation} \label{flr}
Y_i = \alpha + \int_I{X_i(t)\omega(t)}dt + \varepsilon_i,  \; i = 1, \ldots, n,
\end{equation}
where $\alpha$ is the scalar intercept and $\varepsilon_i$ is the error term such that $\varepsilon_i \sim N(0,\sigma^2)$.  $\omega$ is a square integrable coefficient function that relates the predictor process to the response.  The magnitude of $\omega(t)$ indicates the relative importance of the predictor $X_i$ at a given value of $t$.  If $|\omega(t_0)|$ is large, this means that changes in the predictor process at $t_0$ are important in predicting the response.

We note that in model (\ref{flr}), the predictor is presented as though it is a 1-dimensional signal depending on scalar value $t$.  However, $X_i$ could also be a 2- or higher-dimensional functional object such as an image.  In the case of a 2-dimensional predictor image, $t$ corresponds to an ordered pair and $\omega$ is a coefficient image that provides information about the association of various regions of the predictor image with the scalar outcome of interest \citep{Philpaper3, WDSOIR}.

A variety of approaches have been developed for estimating the coefficient function in (\ref{flr}).  Many of these approaches employ functional principal components regression (FPCR) as in \cite{FunctionalGLM, fda2}; and \cite{CaiHall}.  \cite{Reisspaper1} employ both penalization techniques and FPCR or functional partial least squares (FPLS).  Spline-based approaches are also common for estimating $\omega$ \citep{Spline1, Spline2}.  Other ``highly flexible'' approaches include those taken by \cite{MullerYao} who propose a functional additive regression model and \cite{FAME} who extend generalized linear models, generalized additive models, and projection pursuit regression to include functional predictors.  Other more recent developments in estimating $\omega$ rely on combining the use of wavelets and sparse fitting procedures as in \cite{Zhao} and \cite{WDSOIR}.

Although (\ref{flr}) can be appropriate for modeling the relationship between a scalar response and a functional predictor when the association between the response and predictor is the same for all observations, it is inadequate for settings in which the coefficient function differs across subgroups of the observations.  If there are $C$ different associations corresponding to $C$ different coefficient functions then we can think of each observation as coming from one of $C$ distinct subpopulations/components and would need $C$ distinct FLMs to adequately describe the relationship between the response and the predictor.  We are concerned with cases in which subpopulation membership is not observed and will need to be estimated along with the component-specific coefficient functions.  As motivation for a model that accounts for heterogeneous association between a functional predictor and scalar response \cite{ffmrYao} describe a study of the association between the longevity (scalar response) and reproductive trajectory over the first 20 days of life (functional predictor) in female Mediterranean flies.  Their analysis showed evidence of the existence of two distinct subpopulations defined by two different kinds of association between reproductivity and longevity.

Several approaches have been proposed to model scenarios like that mentioned above in both classical regression (with scalar predictors and a scalar response) and scalar-on-function regression.  Some rely on using a clustering algorithm followed by fitting linear models within the estimated clusters as done by \cite{ClustReg} in the case of classical regression and \cite{ClustPLS} in the case of scalar-on-function regression.  An alternative approach, which we adopt here, makes use of the theory of finite mixture regression models.  When the coefficient function is thought to be different for mutually exclusive subgroups, it is natural to model the heterogeneity with a finite mixture of regressions.  Finite mixture models have become an important statistical tool because of their ability to both approximate general distribution functions in a semi-parametric way and account for unobserved heterogeneity \citep{Grun1}.  Furthermore, mixture models can provide insight into previously unknown mechanisms that relate the predictors to the response.

Although the underlying theory of finite mixture regression models and methods for estimating those models have been well-studied when the predictors are scalars \citep{FinMixMod,MedAppFMR}, methods for finite mixture regression remain relatively undeveloped when the predictors are functions.  To our knowledge, \cite{ffmrYao} are the only ones to investigate such an extension.  They refer to the corresponding class of models that use the framework of finite mixture regression to model relationships between functional predictors and a scalar response in the presence of unobserved heterogeneity as functional mixture regression (FMR) models.  In their approach, they first represent each functional predictor in terms of some suitably chosen number of functional principal components and apply standard mixture regression techniques in the new coordinate space.

The FMR model is given by
\begin{equation} \label{ffmr}
Y_i = \alpha_r + \int_I{X_i(t)\omega_r(t)}dt + \varepsilon_i \ \ \ \textrm{if subject \emph{i} belongs to the} \;r\textrm{th group},
\end{equation}
where $C$ is the number of components or distinct subpopulations, $\alpha_r$ is the $r$th component-specific intercept, and $\omega_r$ is the regression function for the $r$th group, $r = 1,\ldots,C$.

In contrast to \cite{ffmrYao}, we propose to take a wavelet-based approach to FMR models. We focus here on the wavelet basis for several reasons.  Wavelets are particularly well suited to handle many types of functional data, especially functional data that contain features on multiple scales.  They have the ability to adequately represent global and local attributes of functions and can handle discontinuities and rapid changes.  

Furthermore a large class of functions can be well represented by a wavelet expansion with relatively few non-zero coefficients.  This is a desirable property from a computational point of view as it aids in achieving the goal of dimension reduction and will be of critical importance when the functional predictors are of very high dimension as is the case when the predictors are 2- or 3-dimensional images.

Once the model is represented in terms of an appropriate wavelet basis, we employ a lasso-type \citep{Tibshiranilasso} $\ell_1$-penalized fitting procedure to estimate the wavelet and scaling coefficients corresponding to the component-specific coefficient functions in the mixture model.

The rest of the paper is organized as follows.  In Section \ref{wavandspec}, we provide a brief discussion of wavelets and the wavelet-based functional linear model followed by specification of the wavelet-based (WB) functional finite mixture regression and adaptive wavelet-based (AWB) functional finite mixture regression models.  In Section \ref{EM}, we outline an EM-type algorithm for fitting WB models.  Section \ref{tuning} discusses the various tuning parameters in the WB and AWB models.  Section \ref{simdata} presents simulation results showing the performance of the WB and AWB methods and an application of our method to a real data set where we investigate the association between fractional anisotropy profiles from the corpus callosum and scores on a test of cognitive functioning in a sample of subjects with multiple sclerosis.  In section \ref{discussion} we conclude with a brief discussion.

\section{Methodology} \label{wavandspec}
\subsection{Wavelets and Wavelet Decomposition}
Wavelet bases are sets of functions that can be employed in representing a wide range of functional data and have the ability to represent localized features of functions in a sparse way.  Comprehensive treatment of wavelets and their applications in statistics can be found in \cite{mybook,NasonR}; and \cite{Vidakovicbook}.  Here, we provide a brief overview of some important aspects of wavelet bases.

In $L^2(\mathbb{R})$, a wavelet basis is generated by two kinds of functions: a father wavelet, $\phi(t)$, and a mother wavelet, $\psi(t)$, with the following properties:
\begin{equation*}
\int\phi(t)dt=1 \textrm{ and } \int\psi(t)dt=0.
\end{equation*}
Father wavelets, also referred to as scaling functions, serve to approximate the function of interest while mother wavelets serve to provide the detail not captured by this approximation.

Any particular wavelet basis consists of translated and dilated versions of its father and mother wavelets given by
\begin{equation*}
\phi_{j,k}(t)=2^{j/2}\phi(2^{j}t-k) \textrm{ and } \psi_{j,k}(t)=2^{j/2}\psi(2^{j}t-k),
\end{equation*}
where the integer $j$ is the dilation index referring to the scale and $k$ is an integer that serves as a translation index.  Larger values of $j$ correspond to scaling and wavelet functions that can provide more localized information about the function of interest.

In practice, the functional predictors and coefficient function associated with the FLM are not considered outside of a given interval.  Without loss of generality, we take that interval to be [0,1].  Wavelet and scaling functions can be adapted via implementation of one of several boundary handling schemes to represent a given function on the unit interval.
Hence if we assume that $\omega \in L^2([0,1])$ then we can represent $\omega$ in the wavelet domain by
\begin{equation} \label{omegawave}
\omega(t)= \displaystyle\sum_{k=0}^{2^{j_0}-1}\beta'_{j_0,k}\phi_{j_0,k}(t)+ \displaystyle\sum_{j=j_0}^{\infty} \displaystyle\sum_{k=0}^{2^j-1}\beta_{j,k}\psi_{j,k}(t),
\end{equation}
where $j_0$ is an integer that determines the number of scaling functions used in the lowest scale representation.  The coefficients $\beta'_{j_0,k}$ and $\beta_{j,k}$ are the corresponding scaling and wavelet coefficients for the functions $\phi_{j_0,k}$ and $\psi_{j,k}$ respectively and are given by
\begin{equation*}
\beta'_{j_0,k}=\int\omega(t)\phi_{j_0,k}(t)dt, \ \ \  \beta_{j,k}=\int\omega(t)\psi_{j,k}(t)dt.
\end{equation*}
Each coefficient provides information about the characteristics of the function $\omega$ at a given scale and location.

We restrict ourselves to orthonormal wavelet families.   By this we mean that $\phi_{j,k}(t)$ and $\psi_{j,k}(t)$ are such that

\begin{singlespace}
\begin{equation*}
\begin{array}{lllcr}
\int\phi_{j',k}(t)\phi_{j',k'}(t)dt & = & \delta_{k,k'} & & \\
\int\psi_{j,k}(t)\phi_{j',k'}(t)dt & = & 0 & \textrm{where} &  \delta_{a,b}=\begin{cases}
                                                                                                            1 & \textrm{ if } a=b \\
                                                                                                            0 & \textrm{ if } a\neq b.
                                                                                                            \end{cases}\\
\int\psi_{j,k}(t)\phi_{j',k'}(t)dt & = & \delta_{j,j'}\delta_{k,k'} & & \\
\end{array}
\end{equation*}
\end{singlespace}
\vspace{0.50cm}
In typical applications, the functional predictors are discretely sampled.  We assume that we observe a dyadic length $(N = 2^J)$ vector of function values $X_i = (X_i(t_1), \ldots, X_i(t_N))^T$ where the arguments $t_1, \ldots, t_N$, are equally spaced and the same for all observations.  To obtain the wavelet and scaling coefficients corresponding to the functional predictors we use the discrete wavelet transform (DWT).  The inverse DWT (IDWT) can be used to reconstruct a vector of functional observations from its corresponding wavelet and scaling coefficients.  Both the DWT and IDWT can be performed using a computationally fast pyramid algorithm \citep{Mallat}.

\subsection{Wavelet Representation of the FLM}
Before moving on to our WB model we first review the WBFLM proposed by \cite{Zhao}.  In the this model, both the coefficient function, $\omega$ and the functional predictor $X_i$ from (\ref{flr}) are expressed in terms of an appropriate wavelet basis.  $\omega(t)$ can be expressed as in (\ref{omegawave}) and $X_i(t)$ can be expressed as
\begin{equation*}
X_i(t)= \displaystyle\sum_{k=0}^{2^{j_0}-1}z'_{i,j_0,k}\phi_{j_0,k}(t) + \displaystyle\sum_{j=j_0}^{\infty} \displaystyle\sum_{k=0}^{2^j-1}z_{i,j,k}\psi_{j,k}(t),
\end{equation*}
where the scaling and wavelet coefficients are given respectively by
\begin{equation*}
z'_{i,j_0,k}=\int X_i(t) \phi_{j_0,k}(t)dt \ \ \textrm{and} \ \ z_{i,j,k}=\int X_i(t) \psi_{j,k}(t)dt.
\end{equation*}

In practice, given $N$ equally-spaced observations of $X_i$, the corresponding wavelet and scaling coefficients can be calculated using the DWT.  These coefficients can be be put into an $(N + 1) \times 1$ vector denoted by $Z_i$ (here we include 1 as the first element of the vector which will correspond to the intercept in the regression model) having the form
\begin{equation} \label{Zi}
Z_i = (1, z'_{i,j_0,0}, \ldots, z'_{i,j_0,k_{j_0}}, z_{i,j_0,0}, \ldots, z_{i,j_0,k_{j_0}}, \ldots, z_{i,J,0}, \ldots, z_{i,J,k_{J}}),
\end{equation}
where $J = \log_2(N) - 1$ and $k_j = 2^j - 1$.

Because of the orthonormality of the wavelet basis, (\ref{flr}) can be simply written as
\begin{equation*}
Y_i =  \alpha + \displaystyle\sum_{k=0}^{2^{j_0}-1}z'_{i,j_0,k}\beta'_{j_0,k} + \displaystyle\sum_{j=j_0}^{J} \displaystyle\sum_{k=0}^{2^j-1}z_{i,j,k}\beta_{j,k} + \varepsilon_i, \ \ \ i = 1,\ldots,n,
\end{equation*}
since all of the cross-product terms comprising the product of the wavelet-domain representations of $X_i$ and $\omega$ integrate to zero \citep{Zhao}, or, in matrix notation:
\begin{equation} \label{matwbmod}
Y = Z\beta + \varepsilon,
\end{equation}
where $Y=(Y_1,\ldots,Y_n)^T$, $Z$ is an $n \times (N + 1)$ matrix with $i$th row $Z_{i}$, $\beta$ is an $(N+1) \times 1$ vector containing the intercept $\alpha$ followed by the coefficients arranged in the same order as the vector $Z_i$, and $\varepsilon = (\varepsilon_1,\ldots,\varepsilon_n)^T$.

Thus we see that once the functions $\omega$ and $X_i$ from (\ref{flr}) have been represented in the wavelet domain, the scaling and wavelet coefficients corresponding to $X_i$, namely the elements of $Z_i$, become the predictors in the transformed space.  In the wavelet representation of (\ref{flr}) there are $N+1$ parameters to estimate (including the intercept).

\subsection{Specification of the WB Functional Mixture Regression Model} \label{specofmodel}
If the pairs of functional predictors and scalar responses come from a heterogeneous population, where the subpopulations (or components) are determined by $C$ distinct associations between the predictors and the response, then there is a unique coefficient function, $\omega_r$, $r = 1,\ldots,C$ corresponding to each subpopulation.  Since, as noted above, the predictors are typically discretely sampled at $N$ points, the model we consider is
\begin{equation*}
Y_i =  \alpha_r + \displaystyle\sum_{k=0}^{2^{j_0}-1}z'_{i,j_0,k}\beta'_{r,j_0,k} + \displaystyle\sum_{j=j_0}^{J} \displaystyle\sum_{k=0}^{2^j-1}z_{i,j,k}\beta_{r,j,k} + \varepsilon_i.
\end{equation*}
Thus, the coefficient functions of interest are given by
\begin{equation*}
\omega_r(t)= \displaystyle\sum_{k=0}^{2^{j_0}-1}\beta'_{r,j_0,k}\phi_{j_0,k}(t)+ \displaystyle\sum_{j=j_0}^{J} \displaystyle\sum_{k=0}^{2^j-1}\beta_{r,j,k}\psi_{j,k}(t)
\end{equation*}
and our goal is to find estimates for the $\beta'_{r,j_0,k}$'s and the $\beta_{r,j,k}$'s.

In this setting, the model of interest is similar to that seen in classical finite mixture regression.  We have that $Y_i|Z_{i} \;\textrm{independent for}\: i = 1,\ldots,n$ and 

\begin{equation}  \label{fmr1}
Y_i|Z_{i} = z \sim \displaystyle\sum_{r=1}^{C}{\pi_r\frac{1}{\sqrt{2\pi}\sigma_r} \textrm{exp}\left(-\frac{(y-z\beta_r)^2}{2\sigma^2_r}\right)} \;\textrm{for} \:i = 1,\ldots,n,
\end{equation}
where $\beta_r$ is the component-specific coefficient vector for component $r$, with the same form as $\beta$ in the model given by (\ref{matwbmod}), $\sigma^{2}_r$ is the corresponding component-specific error variance, and $\pi_r$ is the probability that observation $i$ belongs to component $r$.  Let $\xi = (\beta_1,\ldots,\beta_C,\sigma_1,\ldots,\sigma_C,\pi_1,\ldots,\pi_{C-1}) \in \mathbb{R}^{C(N+1)} \times  \mathbb{R}^{C}_{>0} \times \Pi$ be the  $((N+3)\cdot C - 1)  \times 1$ vector of free parameters to be estimated from (\ref{fmr1}), where $\Pi$ is the space of vectors of the form $(\pi_1,\ldots,\pi_{C-1})$ such that  $\pi_r > 0$  for  $r=1,\ldots,C-1$, $\sum_{r=1}^{C-1}{\pi_r}<1$, and $\pi_C = 1 - \sum_{r=1}^{C-1}{\pi_r}$.

In practice, $X_i$ may be densely sampled and so we may have $N \gg n$.  In this case, maximum likelihood estimation will provide inaccurate and unstable estimates for each $\beta_r$ and consequently poor estimates for each $\omega_r$.  Since wavelets allow for sparse representation of each $\omega_r$, we may assume that most elements of $\beta_r$ are negligible and thus we consider a lasso-type procedure for estimating the $C$ component-specific vectors of wavelet and scaling coefficient values.

\cite{Stadler} proposed an $\ell_1$-penalized mixture regression procedure for model fitting with general high-dimensional predictors.  We make use of this procedure here.  We begin by first reparameterizing model (\ref{fmr1}) using the following:
\begin{equation*}
\varphi_r = \beta_r/\sigma_r, \;\;\;\rho_r = \sigma^{-1}_r, \;\;\; r = 1,\ldots,C.  
\end{equation*}
Based on this new parameterization, model (\ref{fmr1}) can be written as:
\begin{equation} \label{fmr2}
Y_i|Z_{i} = z \sim \displaystyle\sum_{r=1}^{C}{\pi_r\frac{\rho_r}{\sqrt{2\pi}} \textrm{exp}\left(-\frac{1}{2}(\rho_{r}y-z\varphi_r)^2\right)} \;\textrm{for} \:i = 1,\ldots,n.
\end{equation}
There is a one-to-one mapping from $\xi$ in (\ref{fmr1}) to a new parameter vector
\begin{equation*}
\theta =  (\varphi_1,\ldots,\varphi_C,\rho_1,\ldots,\rho_C,\pi_1,\ldots,\pi_{C-1}) \in \mathbb{R}^{C(N+1)} \times  \mathbb{R}^{C}_{>0} \times \Pi.
\end{equation*}
The corresponding log-likelihood for model (\ref{fmr2}) is
\begin{equation} \label{loglik}
\ell(\theta;Y)=\displaystyle\sum_{i=1}^{n}\log\left( \displaystyle\sum_{r=1}^{C}{\pi_r\frac{\rho_r}{\sqrt{2\pi}} \textrm{exp}\left(-\frac{1}{2}(\rho_{r}Y_i-Z_{i}\varphi_r)^2\right)} \right).
\end{equation}

To estimate the parameter vector $\theta$ in model (\ref{fmr2}), we propose to use $\hat{\theta}_\lambda \in \mathbb{R}^{C(N+1)} \times  \mathbb{R}^{C}_{>0} \times \Pi$ that minimizes
\begin{equation} \label{estimator1}
-n^{-1}\ell_{\lambda}(\theta) = -n^{-1}\ell(\theta;Y) + \lambda\displaystyle\sum_{r=1}^{C}\pi_r\Vert\varphi_r\Vert_1,
\end{equation}
where $\Vert\varphi_r\Vert_1$ is the $\ell_1$-norm of the vector $\varphi_r$.  Note that the penalty on each wavelet and scaling component coefficient vector $\varphi_r$ is proportional to the mixing probability $\pi_r$.  Including the mixing proportion in this manner corresponds to the common practice of relating the amount of penalty to the sample size, where, in the context of mixture regression, \cite{KhaliliChen1} note that the ``virtual" sample size from the $r$th component is proportional to $\pi_r$.  Further discussion of the tuning parameters is given in Section \ref{tuning}.

Estimation of $\varphi_r$ and $\rho_r$ rather than the direct estimation of $\beta_r$ and $\sigma_r$ is considered primarily for two reasons.  The reparametrization, along with a lasso-type penalty allows for penalization of both the coefficient vectors of interest and the error variances within each component \citep{Stadler} while maintaining convexity of the optimization problem to be solved.

\subsection{The Adaptive Model}
We also consider an adaptive version of the estimator, $\hat{\theta}_\lambda$, above.  \cite{adaptivelasso} proposed the adaptive lasso which allows for differing weights, adaptively chosen, to be assigned to the coefficients in the $\ell_1$ penalty.  Among other benefits, the use of such weights can serve to provide better performance with respect to variable selection in high dimensional settings.  In the two-stage adaptive lasso procedure, one first finds initial estimates for the coefficients of interest and then uses these estimates (or a transformation of them) as weights for the coefficients in the $\ell_1$ penalty of the lasso procedure.

Our adaptive estimator is denoted by
$\hat{\theta}_{adapt;\lambda}$ and minimizes the following criterion which involves a re-weighted $\ell_1$-norm penalty term:
\begin{equation*}
-n^{-1}\ell_{adapt}(\theta) = -n^{-1}\ell(\theta;Y) + \lambda \displaystyle\sum_{r=1}^{C}\pi_r\displaystyle\sum_{q=2}^{N+1}{w_{r,q}\left|\varphi_{r,q}\right|},
\end{equation*}
where $\varphi_{r,q}$ is the $q$th element of the vector $\varphi_r$ and $w_{r,q} = 1 / \left|\tilde{\varphi}_{r,q}\right|$ where $\tilde{\varphi}_{r,q}$ is the $q$th element of the WB functional mixture regression estimate of component vector $r$.  Since it is possible that some the $\tilde{\varphi}_{r,q}$ values can be zero, which would cause the corresponding $w_{r,q}$ values to be infinite, we add a small constant (0.001) to each $\tilde{\varphi}_{r,q}$ estimate in the fitting algorithm.  Note that we follow \cite{Stadler} and use estimates from the WB functional mixture regression in our weights, but other weighting strategies may also be considered.  

\section{Fitting WB Functional Mixture Models} \label{EM}
Fitting of WB functional mixture regression models is carried out in three main steps: (1) Use the DWT to obtain the wavelet and scaling coefficients corresponding to the functional predictors, (2) use an EM-type algorithm for computing parameter estimates in the wavelet domain, and (3) use the IDWT to obtain estimates of the component-specific coefficient functions in the original domain from the corresponding wavelet and scaling coefficient estimates.  Each step is explained in detail below.

\textbf{Step 1.} Use the DWT to decompose the functional predictors and obtain the corresponding wavelet and scaling coefficients for each predictor.  Here we must choose the wavelet family (e.g.,\ Daubechies' least asymmetric wavelets), number of vanishing moments, lowest level of decomposition ($j_0 \in \left\{0,\ldots,\log_2(N)-1\right\}$), and method for handling the boundaries (e.g.,\ symmetric boundary handling).

The empirical wavelet and scaling coefficients for each predictor curve can be arranged into $(N + 1) \times 1$ vectors, denoted $Z_i$, $i = 1,\ldots,n$, which have the same structure as (\ref{Zi}).  We then form $Z$, an $n \times (N + 1)$ matrix with $i$th row $Z_i$.

\textbf{Step 2.}  We carry out an EM-type algorithm for our setting in a manner similar to that described in \cite{Stadler}.  Consider the unobserved random indicator variable $\Delta_{i,r}$ which designates component membership:
\begin{equation*}
\Delta_{i,r} = \left\{ \renewcommand{\arraystretch}{1}
                       \begin{array}{rl}
                        1 & \mbox{if observation} \  i  \ \mbox{belongs to component} \ r \\
                        0 & \mbox{otherwise}
                       \end{array} \right.
\end{equation*}

Then the expected scaled complete negative log-likelihood and penalized negative log-likelihood are given by
\begin{equation*}
Q(\theta|\theta^{(m)}) = -n^{-1}\mathbb{E}_{\theta'}\left[\ell_{c}\left(\theta; Y, \Delta\right)|Y\right],
\end{equation*}
and
\begin{equation*}
Q_{pen}(\theta|\theta^{(m)}) = Q(\theta|\theta^{(m)}) + \lambda \displaystyle\sum_{r=1}^{C}\pi_r\Vert\varphi_r\Vert_1,
\end{equation*}
respectively.

In the E-step of the fitting procedure, we replace each unobserved group membership indicator, $\Delta_{i,r}$, with its expected value
\begin{equation*}
\hat{\Delta}_{i,r} = \mathbb{E}_{\theta^{(m)}}[\Delta_{i,r}|Y] = \frac{\pi_r^{(m)}\rho_r^{(m)}e^{\frac{1}{2}(\rho_r^{(m)}Y_i-Z_{i}\varphi_r^{(m)})^2}}{\sum_{l=1}^C{\pi_l^{(m)}\rho_l^{(m)}e^{\frac{1}{2}(\rho_l^{(m)}Y_i-Z_{i}\varphi_l^{(m)})^2}}}, \ \ r = 1, \ldots, C, \ i = 1, \ldots n,
\end{equation*}
where $\theta^{(m)} = (\varphi_1^{(m)},\ldots,\varphi_C^{(m)},\rho_1^{(m)},\ldots,\rho_C^{(m)},\pi_1^{(m)},\ldots,\pi_{C-1}^{(m)})$ corresponds to the current vector of values for the parameters at EM-iteration $m$.  Hence we can compute $Q(\theta|\theta^{(m)})$.

The M-step of the fitting procedure is carried out in two stages.  First, we fix each component coefficient vector, $\varphi_r$, at its current value of $\varphi_r^{(m)}$ and improve
\begin{equation*}
-n^{-1}\displaystyle\sum_{i=1}^{n}\sum_{r=1}^{C}\hat{\Delta}_{i,r}\log\left(\pi_r\right) + \displaystyle\sum_{r=1}^{C}\pi_r\Vert\varphi_r\Vert_1
\end{equation*}
with respect to $\left\{\pi; \pi_r > 0 \:\textrm{for} \:r=1,\ldots,C \:\mbox{and} \:\displaystyle\sum_{r=1}^{C}{\pi_r}=1  \right\}$ according to the procedure described in \cite{Stadler}.  This yields an updated estimate of the $\pi$ vector, $\pi^{(m+1)}$.
In the second stage of the M-step, we improve with respect to $\varphi$ and $\rho$.  At this stage, we note that the optimization problem decouples into $C$ distinct convex optimization problems where we seek to minimize each of
\begin{equation*}
-n^{-1}\displaystyle\sum_{i=1}^{n}\hat{\Delta}_{i,r}\log\left(\rho_r^{(m)}\right) +
\frac{1}{2n}\displaystyle\sum_{i=1}^{n}\hat{\Delta}_{i,r}\left(\rho_r^{(m)}Y_i-Z_{i}\varphi_r^{(m)}\right)^2 +
\lambda\pi^{(m+1)}_r\Vert\varphi_r\Vert_1, \ \ r = 1,\ldots,C,
\end{equation*}
with respect to $\varphi_r$ and $\rho_r$.  To solve this set of optimization problems, we implement a coordinate descent algorithm that updates one coordinate at a time while holding the other coordinates fixed at their current values.    The update for $\rho_r$ is given by \citep{Stadler}
\begin{equation*}
\rho_r^{(m+1)} = \frac{\left\langle \tilde{Y}, \tilde{Z} \varphi_r^{(m)}\right\rangle +
                         \sqrt{\left\langle \tilde{Y}, \tilde{Z} \varphi_r^{(m)}\right\rangle^2 +
                         4\left\|\tilde{Y}\right\|^2 \sum_{i = 1}^n\hat{\Delta}_{i,r}}}
                         {2\left\|\tilde{Y}\right\|^2},
\end{equation*}
where $\tilde{Y} = (\sqrt{\hat{\Delta}_{1,r}}Y_1,\ldots,\sqrt{\hat{\Delta}_{n,r}}Y_n)^T$ and $\tilde{Z}$ is a matrix with rows $\sqrt{\hat{\Delta}_{i,r}}Z_i$, $i = 1,\ldots,n$.  Here $\left\langle \cdot,\cdot\right\rangle$ refers to the vector inner product and $\left\|\cdot\right\|$ is the Euclidean norm.   Once the update for $\rho_r$ is computed, we calculate the  update for the unpenalized component-specific intercept using
\begin{equation*}
\varphi_{r,1}^{(m+1)} = \displaystyle \frac{\rho_r^{(m+1)} \sum^n_{i = 1}{\hat{\Delta}_{i,r}Y_i} - \sum_{i = 1}^n\hat{\Delta}_{i,r}\left(\sum^{N+1}_{q = 2}{Z_{i,q}\varphi_{r,q}}\right)}{\sum_{i = 1}^n\hat{\Delta}_{i,r}},
\end{equation*}
where $Z_{i,q}$ is the $q$th element of the vector $Z_{i}$.

The coordinate-wise updates for the remaining $N$ coefficients in each $\varphi_r$ vector are computed as
\begin{equation*}
\varphi_{r,q}^{(m+1)} = \begin{cases}
                          0 & \text{if} \ |S_q| \leq n\lambda(\pi_r^{(m+1)}), \\
                          (n\lambda(\pi_r^{(m+1)}) - S_q)/\left\|\tilde{Z}_{,q}\right\|^2 & \text{if} \
                             S_q > n\lambda(\pi_r^{(m+1)}), \\
                          -(n\lambda(\pi_r^{(m+1)}) + S_q)/\left\|\tilde{Z}_{,q}\right\|^2 & \text{if} \
                             S_q < -n\lambda(\pi_r^{(m+1)}), \\
                          \end{cases}
\end{equation*}
where $\tilde{Z}_ {,q} = \sqrt{\hat{\Delta}_{i,r}}Z_{,q}$ with $Z_{,q}$ being the $q$th column of $Z$ and where we define $S_q$ by
\begin{equation*}
S_q = -\rho_r^{(m+1)}\left\langle \tilde{Z}_ {,q},\tilde{Y}\right\rangle +
        \displaystyle\sum_{s<q}\varphi_{r,s}^{(m+1)}\left\langle
        \tilde{Z}_ {,q},\tilde{Z}_{,s}\right\rangle +
        \displaystyle\sum_{s>q}\varphi_{r,s}^{(m)}\left\langle
        \tilde{Z}_ {,q},\tilde{Z}_{,s}\right\rangle,
\end{equation*}
for $q = 2,\ldots,N+1$.  
The E- and M-steps are iterated until some convergence criteria are satisfied which ensure that the relative improvement in $\ell_{\lambda}(\theta)$ and the relative change in the parameter vector are small.  Specifically, the EM procedure stops when
\begin{equation*}
\frac{\left|\ell_{\lambda}(\theta^{(m+1)})-\ell_{\lambda}(\theta^{(m)})\right|}{1 + \left|\ell_{\lambda}(\theta^{(m+1)})\right|} \leq \tau \ \ \ \  \mbox{and}  \ \ \ \ \underset{q}{\operatorname{max}}\left\{ \frac{\left|\theta_q^{(m+1)} - \theta_q^{(m)}\right|}{1 + \left|\theta_q^{(m+1)}\right|} \right\} \leq \sqrt{\tau},
\end{equation*}
where $\theta_q$ refers to the $q$th element of the parameter vector $\theta$ and we set $\tau = 10^{-6}$.  

\textbf{Step 3.} Use the IDWT to obtain estimates $\hat{\omega}_1,\ldots,\hat{\omega}_C$ from the estimates $\hat{\sigma}_1\hat{\varphi}_1, \ldots, \hat{\sigma}_1\hat{\varphi}_C$ respectively.

The EM procedure discussed in \textbf{Step 2} above requires that we provide initial values for the parameters being estimated.  We use the following scheme for obtaining these initial values.  We first assign a weight to each observation corresponding to each of the $C$ distinct components.  To do this, we randomly assign to each observation $i$ a class, $\kappa$, from the set $\left\{1,\ldots,C\right\}$.  For observation $i$ and its randomly selected  class $\kappa$ we assign $\tilde{\Delta}_{i,\kappa} = 0.9$ and for each of the other classes we assign $\tilde{\Delta}_{i,r} = 0.1$, $r \in \left\{1,\ldots,C\right\}/{\kappa}$.  We then normalize the vector of $\tilde{\Delta}_{i,r}$ values, $r = 1,\ldots,C$ to sum to 1.  Note that this process can be thought of as an initialization of the E-step.  This is followed by updating all of the coordinates involved in the optimizations in the M-step with initial values of $\varphi_{r,q}^{(0)} = 0$, $\rho_r^{(0)} = 2$, and $\pi_r^{(0)} = 1/C$, $r = 1,\ldots,C$, $q = 1, \ldots,N+1$.

To speed up the EM procedure, we restrict ourselves to updating only the non-zero coordinates (active set elements) for  10 out of every 11 iterations of \textbf{Step 2}.  This type of active set algorithm is used in \cite{Meier1,glmnetR}; and \cite{Stadler}.  After 10 iterations on the active set, we expand to consider all coordinates, both the active and non-active, for updating in the 11th iteration.  We obtain a possibly new active set and continue in this manner until the convergence criteria are satisfied.

\section{Tuning Parameters and Their Selection} \label{tuning}
In using WBFMR we need to specify a number of tuning parameters, namely, the number of components, $C$, the lowest level of decomposition, $j_0$, and the penalty parameter, $\lambda$.  The choice of the values for each of these tuning parameters may be based on prior information, otherwise data-driven methods may be employed in their selection.  Below, we discuss each tuning parameter and consider two possible data-driven methods for their selection.

If the number of components is known \emph{a priori} or exploratory data analysis suggests a particular number of components, then $C$ can be specified outright.  However, we often employ the mixture modeling approach when the number of components is unknown or knowledge of component membership is unavailable.

The value of $j_0$ corresponds to the lowest level of decomposition and can range from 0 to $\log_2(N) - 1$.  Since the predictors are sampled at $N$ points, the DWT provides a decomposition that uses a total of $N$ wavelet and scaling functions.  Among this set of $N$ basis functions, $2^{j_0}$ will be scaling functions and $N - 2^{j_0}$ will be wavelet functions.  Hence, setting $j_0$ close or equal to 0 results in using fewer scaling functions to represent large-scale features and more wavelet functions to represent local details of the function of interest.  Conversely, setting $j_0$ close or equal to $\log_2(N) - 1$ results in using more scaling functions and fewer wavelet functions.

The value of $\lambda$ directly determines the role that the penalty function will have in both estimating and selecting variables in the model.  Large values of $\lambda$ force elements of the estimated component coefficient vectors to zero while small values result in many non-zero estimates.    

We will employ two methods for tuning parameter selection.  First we consider selecting the parameters that minimize the cross-validated value
\begin{equation} \label{neg2loglik}
-2\ell(\hat{\theta}_{j_0,\lambda,C}; Y).
\end{equation}
Here, $\ell(\cdot;\cdot)$ denotes the log-likelihood from (\ref{loglik}) and the estimate $\hat{\theta}_{j_0,\lambda,C}$ depends on the values of the tuning parameters as indexed by the subscripts. We will refer to (\ref{neg2loglik}) as the ``predictive loss''. We also consider selecting the parameters that minimize a modified BIC criterion.  We use the modified BIC measure, proposed by \cite{PanShen}, which is given by
\begin{equation} \label{bic}
\textrm{BIC} = -2\ell\left(\hat{\theta}_{j_0,\lambda,C};Y\right) + \log(n)d_e
\end{equation}
where $d_e = (N+3) \cdot C - 1 - q_0$ is the effective number of parameters with $q_0$ being the number of coefficients estimated to be zero in all of the components.  We can compute this value over a grid of candidate values for all or a subset of the turning parameters.

The cross-validation procedure generally puts more emphasis on predictive ability and chooses a model that performs well in this regard.  On the other hand, BIC focuses more on finding the ``true'' model and often chooses a simpler one.  Compared to BIC, cross-validation is computationally demanding and can be prohibitive for large and/or high-dimensional data.

\section{Simulations and Application} \label{simdata}
We present simulation results that demonstrate various aspects of the WBFMR and AWBFMR procedures and that draw comparisons to a functional principal components-based (FPC) method similar to that proposed by \cite{ffmrYao}.  For each simulation discussed below, we generated observations consisting of a discretely sampled one-dimensional functional predictor signal, $X_i$, and a scalar response, $Y_i$ whose association with $X_i$ depends on some known group membership.

\begin{figure}[ht]
\centering
\includegraphics[clip = true, totalheight = 0.22 \textheight, angle=0]{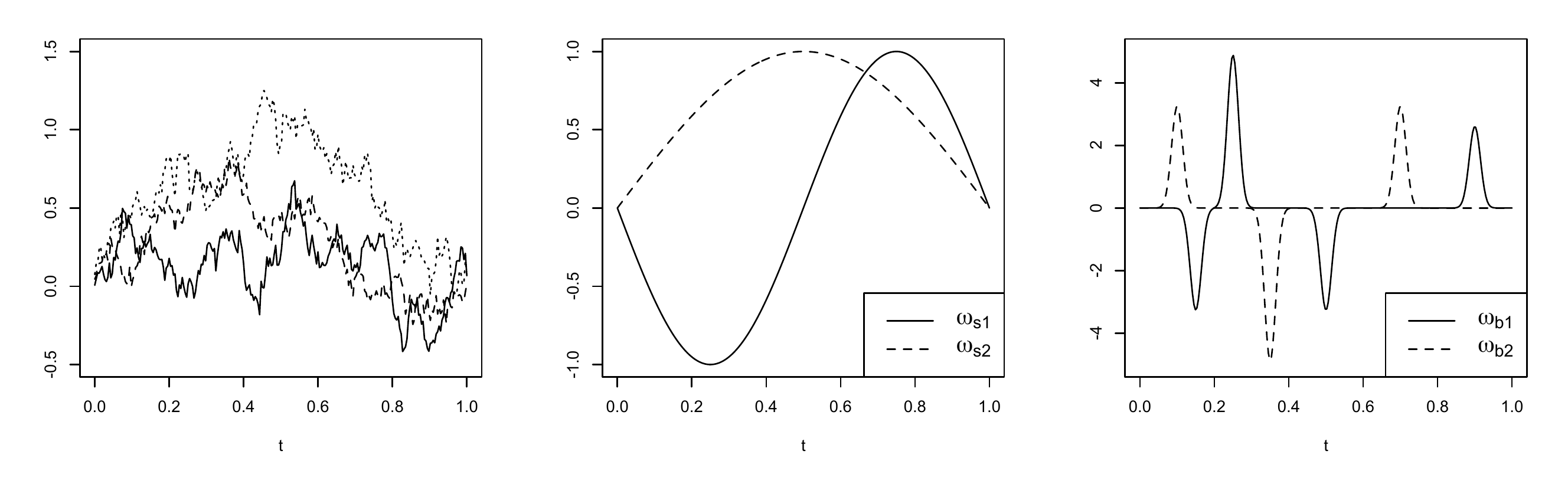}
\caption{Left: Sample of three predictor curves ($N = 256$). Center: Component coefficient functions in smooth setting ($\omega_{s1}$ and $\omega_{s2}$).  Right: Component coefficient functions in bumpy setting ($\omega_{b1}$ and $\omega_{b2}$).}
\label{fig:compcoeff}
\end{figure}

Each functional predictor is a Brownian bridge stochastic process for $t \in (0,1)$ with an expected value of 0, covariance given by $cov(X_i(t),X_i(s)) = s(1-t)$ for $s < t$, and with $X_i(0) = X_i(1) = 0$.  We consider various sampling densities for the functional predictors.  Specifically, we consider data sets where the functional predictors are sampled at $N=$ 64, 128, 256, or 512 equally-spaced points.  A sample of three of these predictors are given in the left panel of Figure \ref{fig:compcoeff}.

The scalar outcomes corresponding to each functional predictor were generated using two distinct settings for the component-specific coefficient functions.  The first pair of component-specific coefficient functions are given by $\omega_{s1}(t)= -sin(2\pi t)$ and $\omega_{s2}(t)= sin(\pi t)$.  The second pair of component-specific coefficient functions are given by $\omega_{b1}(t)=-3.257e^{-a(t-0.15)^2} + 4.886e^{-a(t-0.25)^2} - 3.257e^{-a(t-0.5)^2} + 2.606e^{-a(t-0.9)^2}$ and $\omega_{b2}(t)= 3.257e^{-a(t-0.1)^2} - 4.886e^{-a(t-0.35)^2} + 3.257e^{-a(t-0.7)^2}$ where $a = 20000/9$.  The middle panel of Figure \ref{fig:compcoeff} shows $\omega_{s1}$ and $\omega_{s2}$ which we will refer to as the ``smooth'' functions while the right panel shows $\omega_{b1}$ and $\omega_{b2}$ which we will refer to as the ``bumpy'' functions.

In addition to considering different component-specific coefficient settings, we also consider different signal-to-noise ratio settings.  Equal proportions of observations were generated in each component with $\sigma_1 = \sigma_2$ and the discrete approximation to
$R^2 = \sum_{r = 1}^2\pi_rvar(X\omega_r)/ \\
 \sum_{r = 1}^2\pi_r(var(X\omega_r)+\sigma_r^2) $
takes on the desired value in a given setting.  We consider settings with $R^2$ values of 0.9, 0.7, and 0.5 corresponding to ``high'', ``medium'', and ``low'' signal-to-noise ratios respectively.

We employ Daubechies' least asymmetric wavelets with eight vanishing moments in all simulations.  The \verb+WaveThresh+ package in \verb+R+ \citep{wavethreshSoft} is used to perform the DWT and IDWT with the periodic boundary handling option.

\subsection{Simulation 1: Comparison of FMR Methods}
In the first set of simulations we compare the wavelet-based (WB), adaptive wavelet-based (AWB), and functional principal components-based (FPC) mixture regression methods in various combinations of the settings mentioned above. In all, we consider 24 different settings: two types of component coefficient functions (smooth or bumpy), three possible $R^2$ values (0.9, 0.7, or 0.5), and four possible sampling densities ($N$ = 64, 128, 256, or 512).

For a given simulation run, we generate a training set, a validation set, and a test set all from the same setting (i.e., \ type of component coefficient functions, $R^2$ value, sampling density, and equal proportion of observations from each component).  Each set is made up of 100 observation pairs consisting of a functional predictor and its corresponding scalar response.  The training set is used to fit a model for each combination of the tuning parameters.  The validation set is then used to select the combination of tuning parameters that minimizes (\ref{neg2loglik}) among all combinations of tuning parameters.  Finally, the model estimated from the validation set is applied to the test set and the corresponding predictive loss is computed.  We repeat this procedure 100 times for each setting.

In this first set of simulations, we treat the number of components as known, i.e., $C = 2$.  For the wavelet-based methods, we fix the lowest level of decomposition to $j_0 = 0$ in the smooth setting and to $j_0 = 5$ in the bumpy setting.  In extensive prior simulations (not shown here), these decomposition levels tended to consistently minimize the predictive loss for each of the settings that we consider here.  Additional support for these choices is provided by the results in Simulation 2 where we note that the plots in Figure \ref{fig:sim2plot} show little difference between the distribution of the losses when $j_0$ is set to the values that we selected and the distributions when $j_0$ is allowed to be chosen by either cross-validation or BIC.  To choose an optimal value for $\lambda$ in a given setting, we first fit the model based on the training data for each $\lambda$ in a grid of 100 values then chose the $\lambda$ value that minimizes (\ref{neg2loglik}) in the corresponding validation set.  The predictive loss is then obtained by using the model fit on the training set to compute (\ref{neg2loglik}) from the corresponding test data.  For the adaptive wavelet-based method, we use the estimated component-specific wavelet and scaling coefficients from the corresponding wavelet-based model as the initial estimates.

For the functional principal components-based procedure, based on the procedure proposed in \cite{ffmrYao}, the tuning parameters consisted of the number of order four B-spline basis functions used in representing the predictor signals and the number of principal components to serve as the predictors in the FMR model.  Following \cite{ffmrYao}, we choose to use the minimum number of principal components that account for at least 90\% of the variation in the predictor signals.  The optimal set of tuning parameters was selected by first fitting a model for each combination of number of B-spline basis functions and number of principal components using the training data and then picking the pair that minimized (\ref{neg2loglik}) in the corresponding validation set.  The fitted model was then applied to the corresponding test data and the predictive loss was obtained.  We use the \verb+FlexMix+ packge \citep{FlexMix} in \verb+R+ to fit the functional principal components-based models.

\begin{sidewaysfigure}[!htbp]
\centering
\includegraphics[clip = true, totalheight = 0.57 \textheight, angle=0]{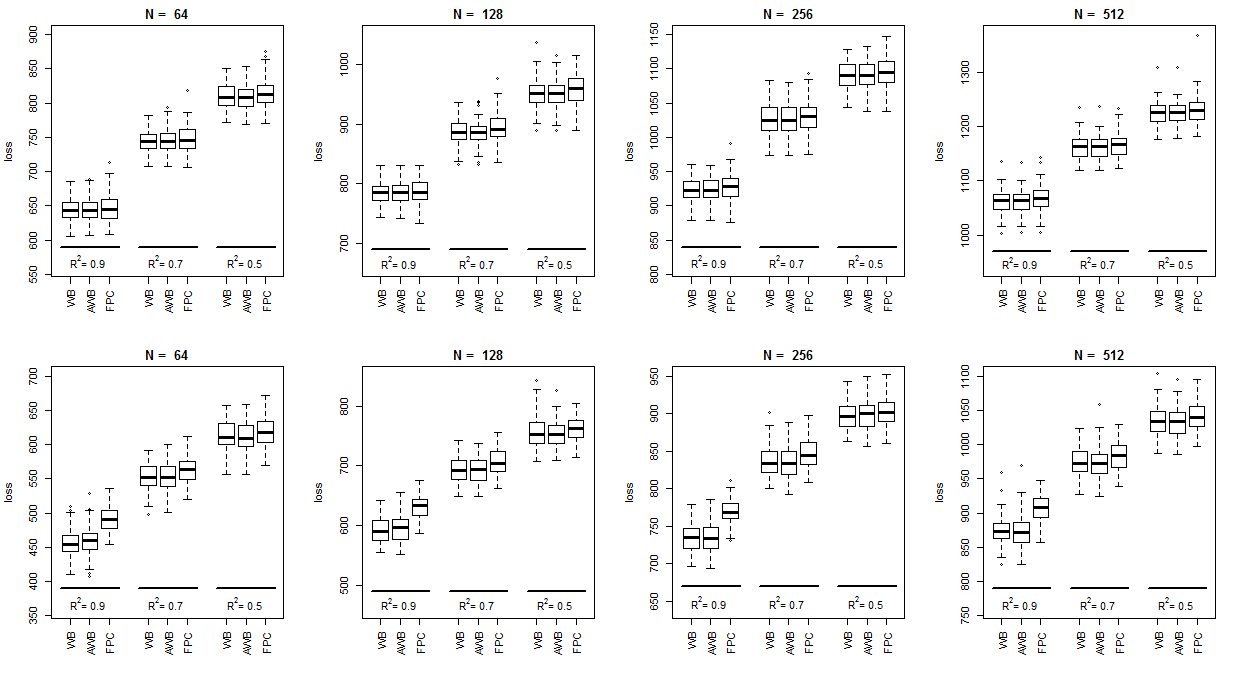}
\caption{Log-likelihood test loss comparing Wavelet-Based (WB), Adaptive Wavelet-Based (AWB), and Functional Principal Components-Based (FPC) methods at each setting; leftmost boxes in each side-by-side plot correspond to $R^2 = 0.9$, middle boxes correspond to $R^2 = 0.7$, rightmost boxes correspond to $R^2 = 0.5$.  Top row corresponds to the smooth setting and bottom row corresponds to the bumpy setting.}
\label{fig:allloss}
\end{sidewaysfigure}

We first consider how the three methods compare with respect to predictive loss based on the test sets.  The boxplots in Figure \ref{fig:allloss} show the predictive loss in the test set for the 100 simulation runs for each of the three methods at the various settings.  Lower loss values are preferred.  In the smooth setting (top row), we note that the wavelet-based and adaptive wavelet-based methods perform comparably to the functional principal components-based method while in the bumpy setting (bottom row), the wavelet-based methods appear to do better, especially for higher values of $R^2$.

\begin{figure}[!htbp]
\centering
\includegraphics[clip = true, totalheight = 0.35 \textheight, angle=0]{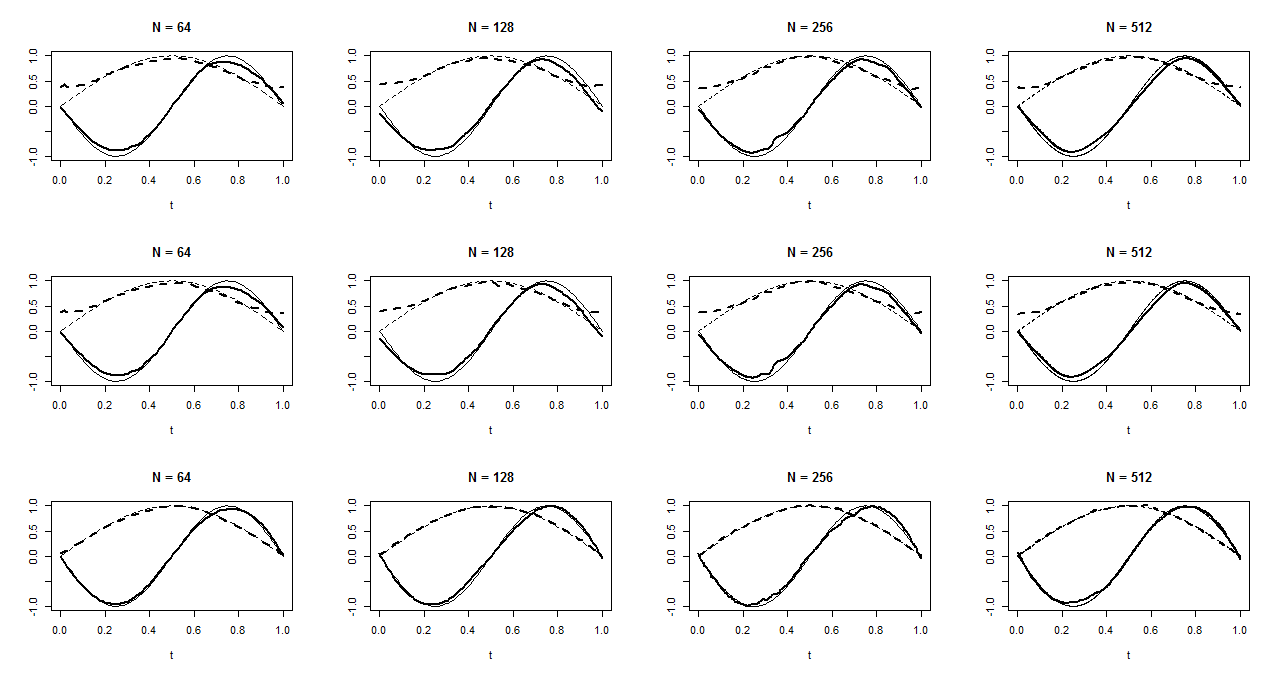}
\caption{Smooth true component coefficient functions; $R^2 = 0.9$; solid and dashed thin curves correspond to the truth; solid dashed thick curves correspond to the point-wise mean estimated component coefficient functions; top row depicts WB method; middle row depicts AWB method; bottom row depicts FPC method.}
\label{fig:sim1.estperfsmooth1}

\includegraphics[clip = true, totalheight = 0.35 \textheight, angle=0]{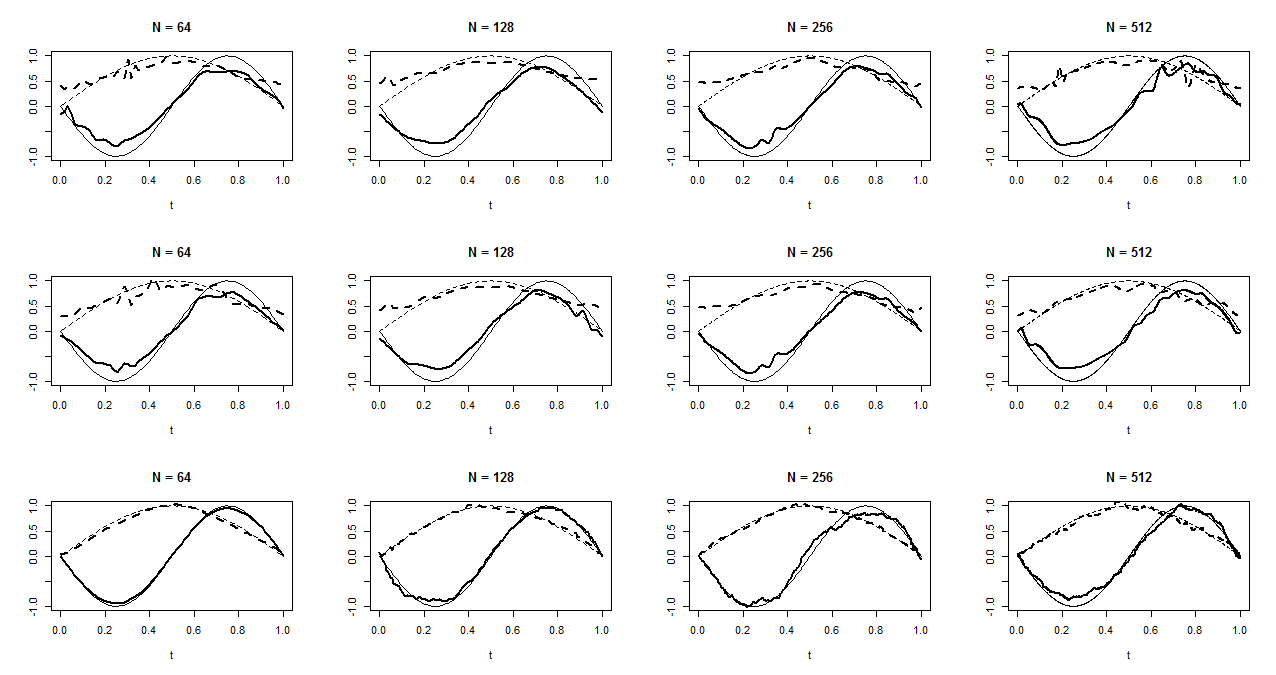}
\caption{Smooth true component coefficient functions; $R^2 = 0.7$; solid and dashed thin curves correspond to the truth; solid dashed thick curves correspond to the point-wise mean estimated component coefficient functions; top row depicts WB method; middle row depicts AWB method; bottom row depicts FPC method.}
\label{fig:sim1.estperfsmooth2}
\end{figure}

\begin{figure}[!htbp]
\centering
\includegraphics[clip = true, totalheight = 0.35 \textheight, angle=0]{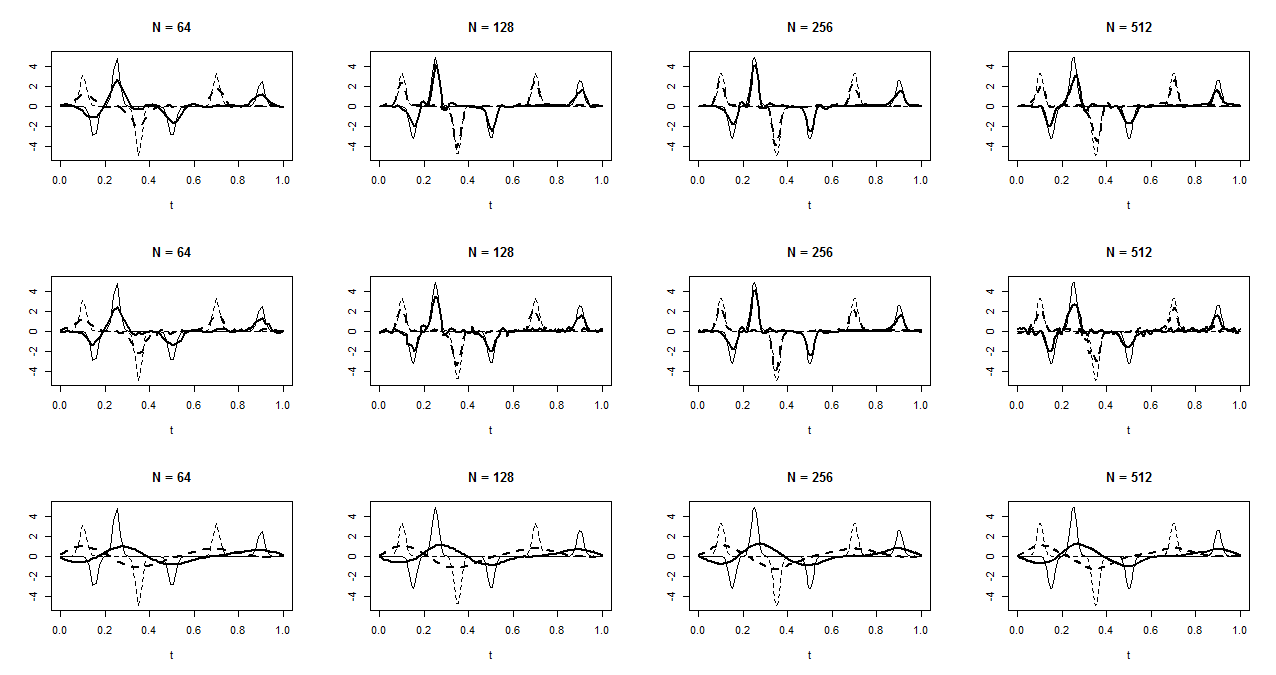}
\caption{Bumpy true component coefficient functions; $R^2 = 0.9$; solid and dashed thin curves correspond to the truth; solid dashed thick curves correspond to the point-wise mean estimated component coefficient functions; top row depicts WB method; middle row depicts AWB method; bottom row depicts FPC method.}
\label{fig:sim1.estperfbump1}

\includegraphics[clip = true, totalheight = 0.35 \textheight, angle=0]{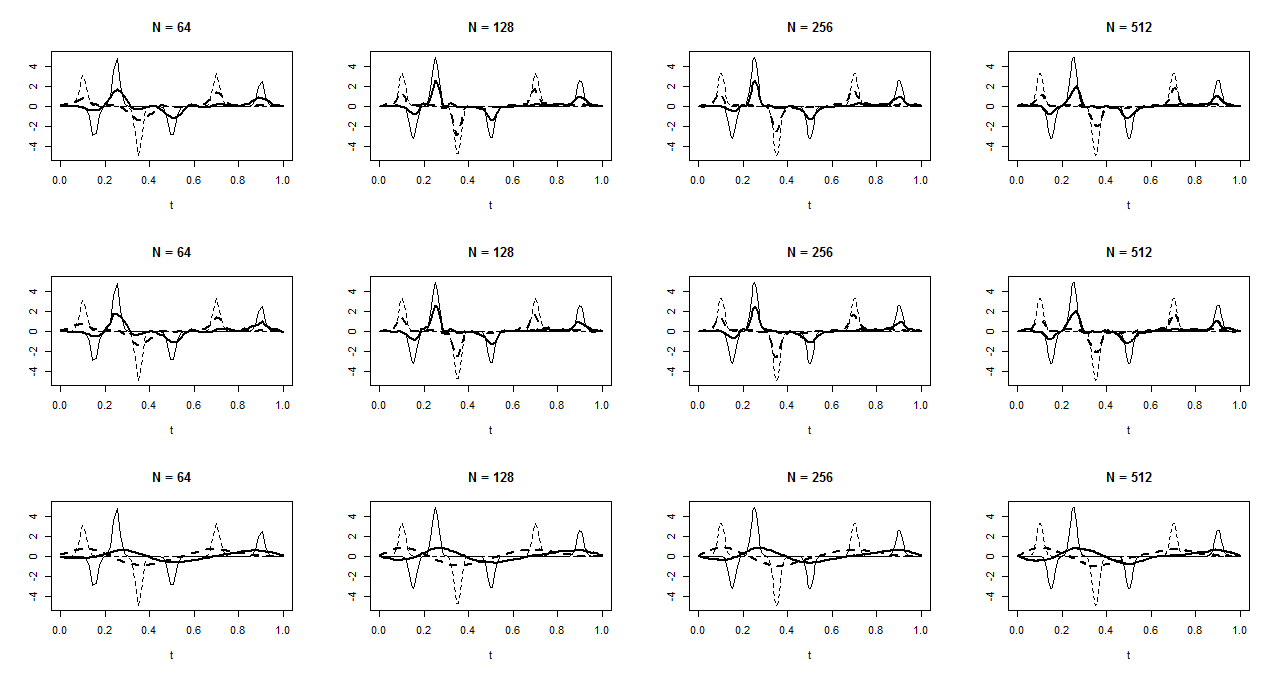}
\caption{Bumpy true component coefficient functions; $R^2 = 0.7$; solid and dashed thin curves correspond to the truth; solid dashed thick curves correspond to the point-wise mean estimated component coefficient functions; top row depicts WB method; middle row depicts AWB method; bottom row depicts FPC method.}
\label{fig:sim1.estperfbump2}
\end{figure}

Estimation performance is illustrated in Figures \ref{fig:sim1.estperfsmooth1}-\ref{fig:sim1.estperfbump2}.  The solid and dashed thick curves correspond to the point-wise mean estimated component coefficient functions over the 100 simulation runs at the specified setting.  The solid and dashed thin curves correspond to the true component coefficient functions used to generate the scalar responses.  Figures \ref{fig:sim1.estperfsmooth1} and \ref{fig:sim1.estperfsmooth2} show the average estimation performance of the three methods in the smooth setting for $R^2$ = 0.9 and 0.7 respectively.  (Performance for $R^2 = 0.5$ is not shown but is similar to that of $R^2 = 0.7$).  In all settings for which the true component coefficient functions are smooth, we note that the functional principal components-based method appears to do best while the wavelet-based methods perform similarly well when the functional predictors are densely sampled.  Substantial gains in estimation performance by the wavelet-based methods are evident in Figures \ref{fig:sim1.estperfbump1} and \ref{fig:sim1.estperfbump2} which show the average estimation performance of the three methods in the bumpy setting for $R^2$ = 0.9 and 0.7 respectively.  (Again, performance for $R^2 = 0.5$ is not shown but is similar to that of $R^2 = 0.7$).  We note that the wavelet-based methods do very well in capturing the local features of the component coefficient functions and in estimating regions where there is no association between the functional predictors and the response while the functional principal components-based method struggles with both of these tasks.

It is interesting to note that the wavelet-based and adaptive wavelet-based methods perform nearly identically in the simulations discussed above.  Comparison (not shown here) of the wavelet and scaling coefficient estimates given by the wavelet-based and corresponding adaptive wavelet-based methods shows that the adaptive version is performing additional variable selection and producing different estimates from those given by the non-adaptive procedure, but these changes do not yield substantial gains in reducing either predictive loss or estimation error.

\subsection{Simulation 2: Tuning Parameter Selection Methods}

In the second set of simulations, we investigate selection methods for the tuning parameters in the wavelet-based model.  We compare selection based on minimizing the 5-fold cross-validated log-likelihood loss to that based on minimizing the modified BIC criteria given in (\ref{bic}).  We consider three different scenarios for tuning parameter selection:
\begin{itemize}[leftmargin = 2.8cm]
\item[Scenario 1.] Set $C = 2$ and $j_0$ = 0 (smooth setting) or 5 (bumpy setting);  select $\lambda$.
\item[Scenario 2.] Set $j_0$ = 0 (smooth setting) or 5 (bumpy setting); select $C \in\{1,2,3\}$ and $\lambda$.
\item[Scenario 3.] Set $C = 2$; select $j_0 \in \{0,\ldots,\log_2(N)-1\}$ and $\lambda$.
\end{itemize}

We restrict ourselves to a subset of four of the 24 settings from the first group of simulations discussed above.  Specifically, we compare the three tuning parameter selection scenarios in the smooth and bumpy settings when the sampling density of the functional predictors is either 128 or 256.  In all four settings we have $R^2 = 0.9$.  For each of 100 simulation runs at each setting, the training set was used to determine the optimal tuning parameters that either minimized the 5-fold cross-validated predictive log-likelihood loss or minimized the modified BIC criteria.  The corresponding test set was used to estimate the test loss in each scenario for both selection methods.

\begin{figure}[!htbp]
\centering
\includegraphics[clip = true, totalheight = 0.6 \textheight, angle=0]{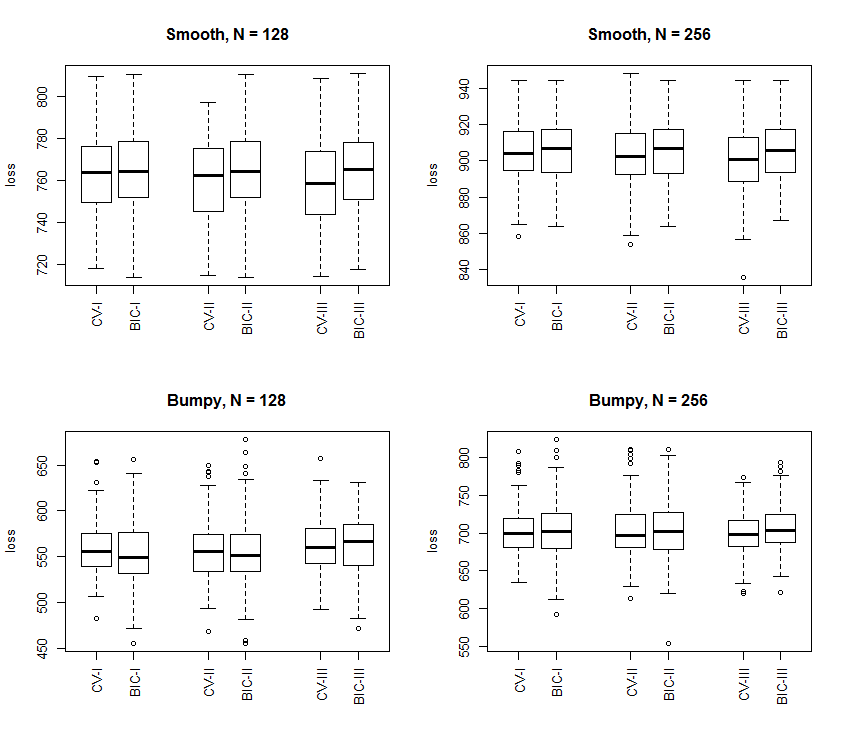}
\caption{Log-likelihood test loss for different tuning parameter selection scenarios ($R^2 = 0.9$); CV-I and BIC-I (leftmost: selection methods for scenario 1), CV-II and BIC-II (middle: selection methods for scenario 2), CV-III and BIC-III (rightmost: selection methods for scenario 3).}
\label{fig:sim2plot}
\end{figure}

The boxplots showing these log-likelihood loss values for the test sets are provided in Figure \ref{fig:sim2plot}.  The three tuning parameter selection scenarios appear to be comparable with respect to predictive log-likelihood loss across the four settings.

In Scenario 2, we allowed the data to select the number of components, $C$.  Table \ref{tab:tablesim2} shows the proportions of simulation runs at each of the four settings for which the number of components was chosen to be 1, 2, or 3.  The table suggests that, relative to using the modified BIC for selecting the number of components, 5-fold cross validation has greater tendency to overfit by estimating more components than truly exist.

\begin{center}
\begin{table}
\centering
\caption{Proportion of simulation runs at each setting such that the indicated number of components ($C$) is chosen by either 5-fold cross-validation or modified BIC.}
\begin{tabular}{|c|c|c|c|c|}
\hline
 & & $C = 1$ & $C = 2$ & $C = 3$ \\
\hline \hline
 Smooth     & CV  & 0.00 & 0.62 & 0.38 \\
 $N = 128 $ & BIC & 0.00 & 1.00 & 0.00 \\
\hline
 Smooth     & CV  & 0.00 & 0.59 & 0.41 \\
 $N = 256 $ & BIC & 0.00 & 1.00 & 0.00 \\
\hline
 Bumpy      & CV  & 0.04 & 0.76 & 0.20 \\
 $N = 128 $ & BIC & 0.05 & 0.94 & 0.01 \\
\hline
 Bumpy      & CV  & 0.07 & 0.72 & 0.21 \\
 $N = 256 $ & BIC & 0.05 & 0.94 & 0.01 \\
\hline
\end{tabular}
\label{tab:tablesim2}
\end{table}
\end{center}

\subsection{Application to DTI Data for Subjects with Multiple Sclerosis}
We now analyze data from a diffusion tensor imaging (DTI) study, discussed in \cite{GoldsmithLongitudinal}, using our wavelet-based functional mixture regression approach.  The data are from a longitudinal study investigating the cerebral white matter tracts of subjects with multiple sclerosis (MS) recruited from an outpatient neurology clinic and healthy controls who were recruited from the community.  Here we focus on the baseline observations for the 100 MS subjects.  In particular, we are interested in the relationship between the fractional anisotropy profile (FAP) from the corpus callosum (functional predictor) and the Paced Auditory Serial Addition Test (PASAT) score (scalar response).

\begin{figure}[!htbp]
\centering
\includegraphics[clip = true, totalheight = 0.45 \textheight, angle=0]{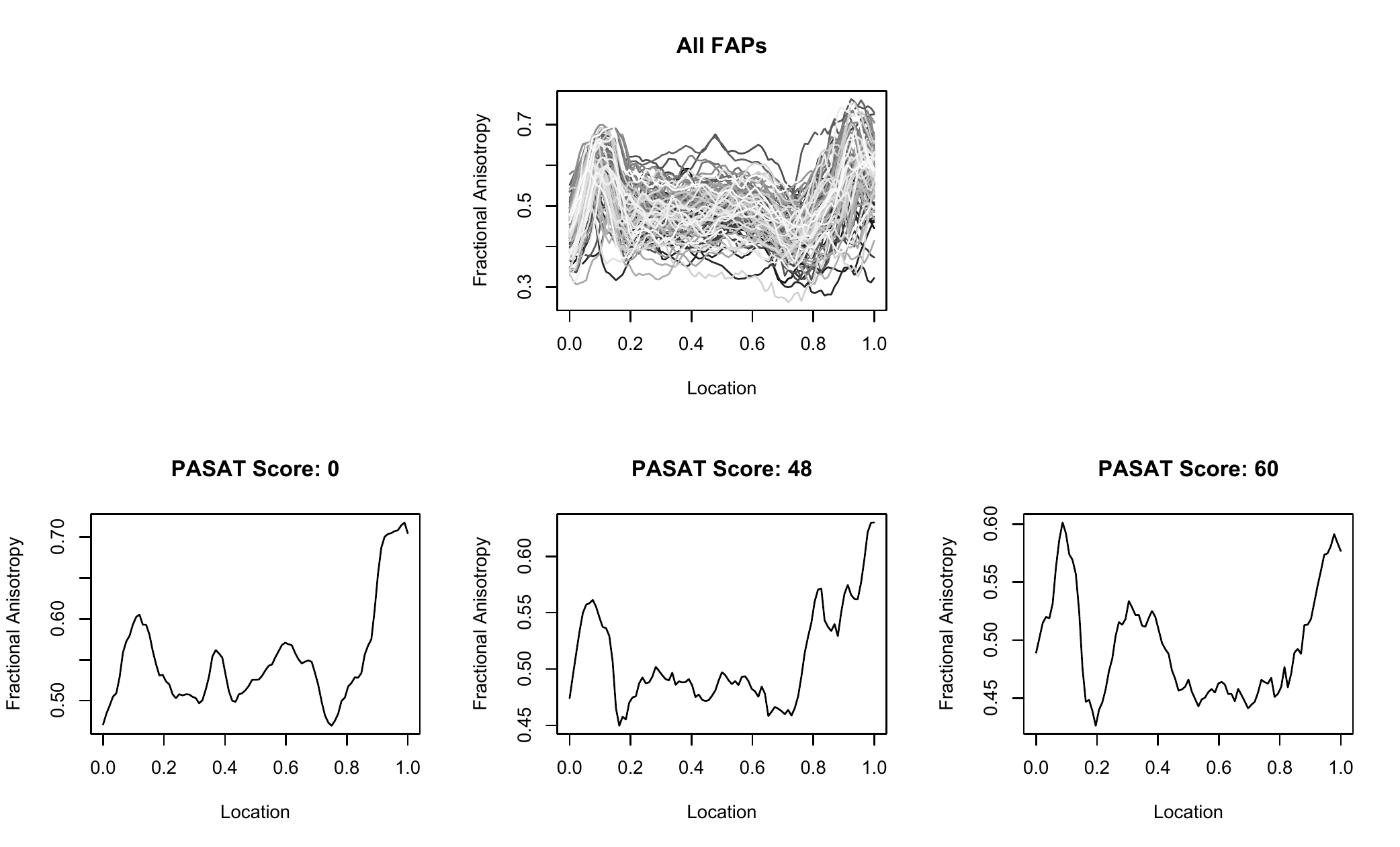}
\caption{FAPs for all subjects (top), subject with lowest PASAT score (bottom left), subject with median PASAT score (bottom center), and subject with highest (bottom right) PASAT score.}
\label{fig:FATP}
\end{figure}

The PASAT is an assessment tool that measures a subject's cognitive ability with respect to auditory information processing speed and flexibility and also provides information on calculation ability \citep{PASAT}.  The PASAT score is the number of correct answers out of 60 questions and thus ranges from 0 to 60.  Lower scores are generally taken to indicate some level of dysfunction. The functional predictor of interest is the FAP from the corpus callosum which is derived from DTI, a magnetic resonance imaging modality that is commonly used to track the diffusion of water in biological tissue.  The FAP is a continuous summary of water diffusivity that is parametrized by the arc length along a curve.  The tract profiles are estimated via an automated tract-probability-mapping scheme described in \cite{DTIdata}.  In the data set, the FAP predictors are recorded at 93 locations along the corpus callosum.  In our analysis, we linearly interpolate the FAP curves at 128 equally spaced points before projecting them onto a wavelet basis.  We used data from 99 of the 100 MS subjects since one subject had missing FAP values at several locations along the tract.  Figure \ref{fig:FATP} shows the FAPs for all 99 MS subjects that we considered as well as those for three subjects with the lowest, median, and highest PASAT scores.

We were interested in conducting an analysis that inspects whether the regression relationship between corpus callosum FAPs and PASAT scores varies due to some unknown mechanism.  In the top plot of Figure \ref{fig:FATP}, we note that there is no obvious grouping in the FAP curves.

\begin{figure}[!htbp]
\centering
\includegraphics[clip = true, totalheight = 0.45 \textheight, angle=0]{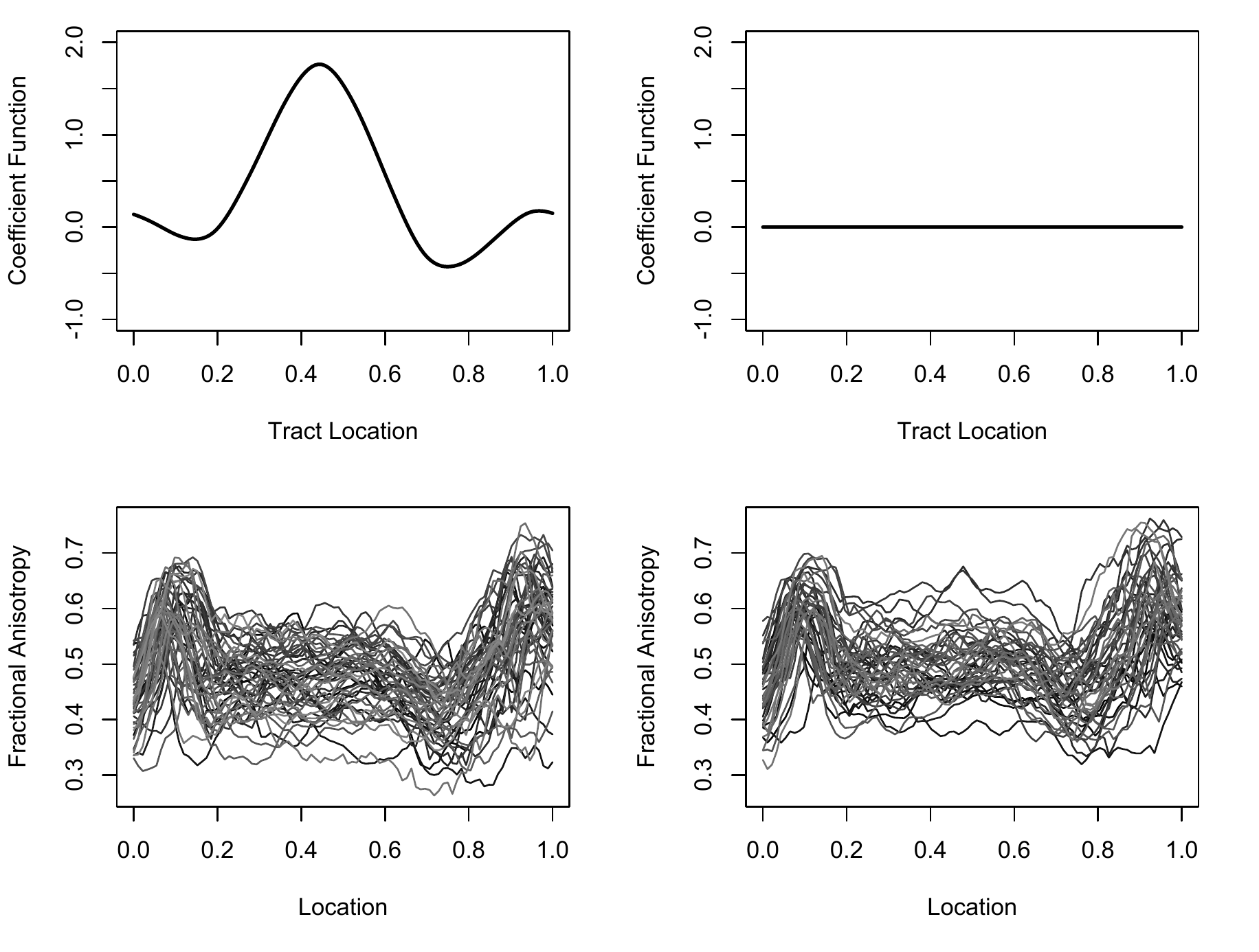}
\caption{Top panels: estimated coefficient functions $\hat{\omega}_1$ (left) and  $\hat{\omega}_2$ (right) of the two groups determined by the WB functional mixture regression approach.  Bottom panels: FAP curves for MS subjects that correspond to $\hat{\omega}_1$ (left) and for those that correspond to $\hat{\omega}_2$ (right).}
\label{fig:FATPcoeffs}

\includegraphics[clip = true, totalheight = 0.42 \textheight, angle=0]{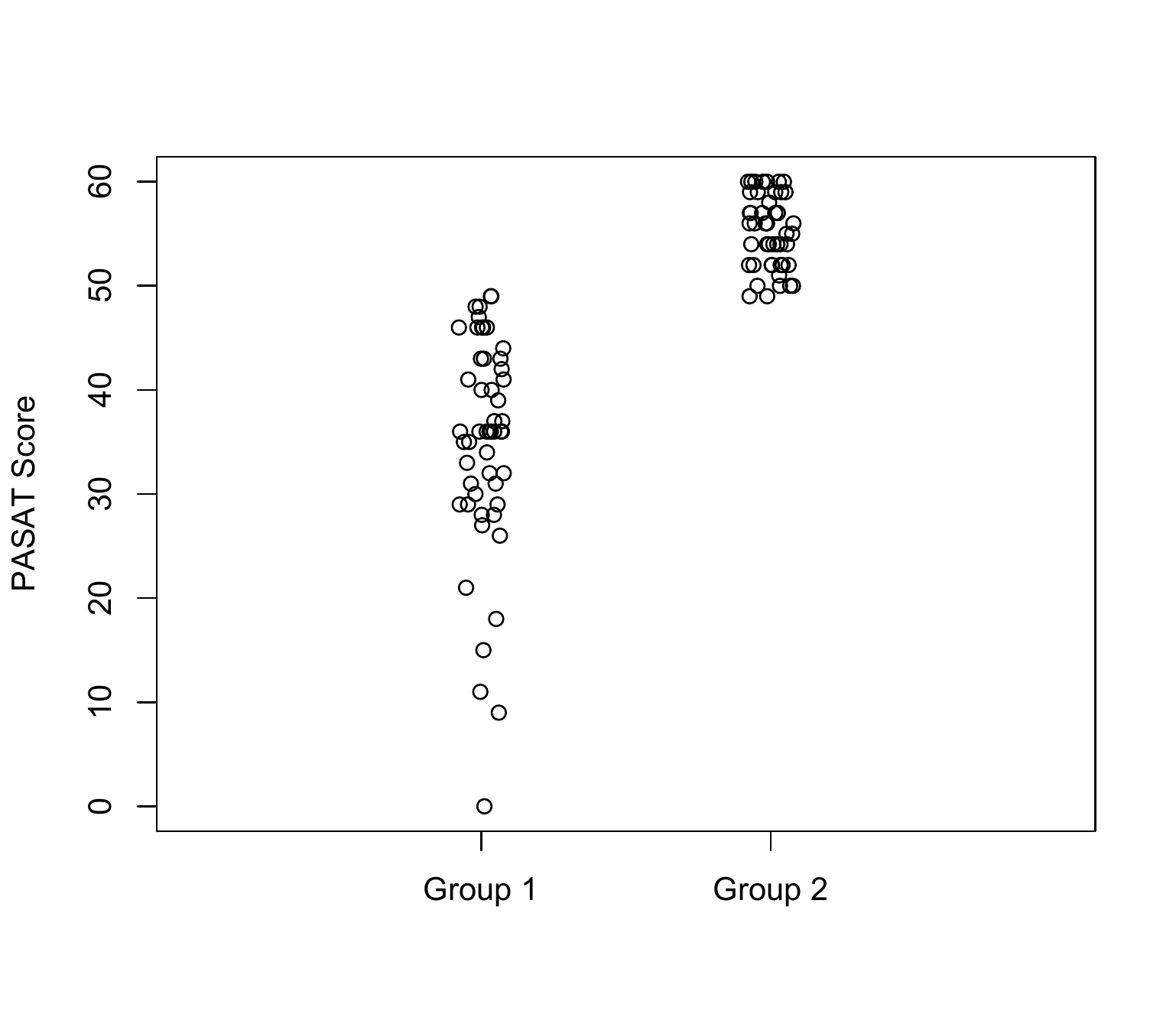}
\caption{PASAT Scores for those in Group 1 (corresponding to $\hat{\omega}_1$) and in Group 2 (corresponding to $\hat{\omega}_2$).}
\label{fig:GroupPASATs}
\end{figure}

We apply our WB functional mixture regression approach in which we used the BIC from (\ref{bic}) to select the optimal tuning parameters.  This approach suggests that there are two distinct groups with different coefficient functions describing the association between corpus callosum FAP and PASAT score.  Figure \ref{fig:FATPcoeffs} shows the estimated coefficient functions, $\hat{\omega}_1$ and $\hat{\omega}_2$, for each of the two groups.  For illustration, Figure \ref{fig:FATPcoeffs} also shows the FAPs that belong to the groups associated with those functions.  To determine which group a subject's FAP belongs to, we use the estimated group membership indicators from the last iteration of the EM algorithm.  The indicator with the the highest value (i.e.,\ $\mbox{max}\left\{\hat{\Delta}_{i,1},\hat{\Delta}_{i,2}\right\}$ for subject $i$) was taken to correspond to the group from which the observation came.  Using this assignment method, there are 52 subjects belonging to Group 1 and 47 belonging to Group 2. 

From Figure \ref{fig:FATPcoeffs} we note that the estimated coefficient function corresponding to Group 2 is identically zero at all locations along the profile suggesting no association between FAP and PASAT score among MS subjects belonging to this group whereas the estimated coefficient function for Group 1 suggests that higher fractional anisotropy values between profile locations of about 0.2 and 0.7 are associated with higher PASAT scores while higher values between profile locations of about 0.7 and 0.9 are associated with lower PASAT scores for those MS subjects belonging to Group 1.

Figure \ref{fig:GroupPASATs} shows the PASAT scores corresponding to the two groups.  This plot illustrates a distinctive split between the two groups with respect to PASAT score. Overall the model may suggest that, among MS subjects with better cognitive function, there is no association between corpus callosum FAP and PASAT score whereas among those with worse cognitive function, fractional anisotropy values in the middle region of the tract can discriminate among the PASAT scores and that greater fractional anisotropy corresponds to higher scores.

With respect to the the properties that characterize the estimated components, the application of our method to the DTI data resulted in findings that are similar to those found in an application of the FPC-based FMR method used in \cite{ffmrYao}.  Here we are referring to the analysis of the association between early reproductivity (functional predictor) and longevity (scalar response) in Mediterranean fruit flies.  As in our application, \cite{ffmrYao} found that their approach suggested there were two groups of flies corresponding to two different regression structures that characterized the association between early fertility and longevity.  Furthermore, upon examining the distribution of the response within each group, they found that one group tended to consist of flies with greater longevity while the other group consisted of flies with shorter longevity; similar to how the estimated groups in our DTI example show a distinction by higher and lower PASAT score.

Finally, we compare the chosen wavelet-based function mixture regression model to the wavelet based FLM (selected to minimize BIC) with respect to leave-one-out cross-validated relative prediction errors $\mbox{CVRPE} = \sum^n_{i = 1}\left(Y_i-\hat{Y}^{(-i)}_i\right)^2/\sum^n_{i = 1}Y^2_i$ where $\hat{Y}^{(-i)}_i$ is the predicted PASAT score for subject $i$ from a model fit on data with subject $i$ removed.  To determine which estimated coefficient function to use to obtain the predicted PASAT score for subject $i$, we use the following ad hoc method similar to that used in \cite{ffmrYao}: if the observed PASAT score $Y_i$ is less than 50 then we use the coefficient function that is not identically zero at each profile location and if $Y_i$ is 50 or larger then we use the zero function.  For our model with 2 groups, the CVRPE is 0.0315 and 0.0723 for the wavelet based FLM.

\section{Discussion} \label{discussion}
In this article we present a general wavelet-based approach to functional mixture regression which is appropriate to use when modeling the association between a continuous scalar response and a functional predictor where the association is not homogeneous across the population.  We provide a fitting algorithm and demonstrate some properties of the corresponding estimators using simulations.  When compared with a functional principal components-based approach to functional mixture regression, evidence suggests that our method performs better with respect to prediction and estimation accuracy when the component coefficient functions defining the association between the predictors and responses possess relatively small scale features.

\cite{Zhao} note that there are many factors that may be important to the performance of a wavelet-based approach like the one we present here.  For one, selection of a particular wavelet basis for the DWT has an impact on the sparsity of the representation of the functional predictor.  As in \cite{Zhao}, we chose to use a wavelet basis from the Daubechies family in our simulations.  This family has good localizing properties in both the temporal and frequency domains.  Of course, selection from other families is possible and it may be of interest to compare performance of our method when basis functions from different families are chosen.

Another factor that plays an important role in the performance of our method is tuning parameter selection.  We looked at two criteria for selecting tuning parameters: minimizing the 5-fold cross-validated predictive loss and minimizing a modified BIC value.  We found that both methods were generally comparable with the BIC method perhaps slightly under-performing with respect to predictive loss.  However, simulations showed that BIC tended to select the correct number of components more often and 5-fold cross-validation tended to overfit.

We noted in Section \ref{specofmodel} that it is common to relate the amount of penalty on the covariates to the sample size as is done in (\ref{estimator1}) by including $\pi_r$ in the penalty function.  In their $\ell_1$-penalized mixture approach, \cite{Stadler} suggest including an additional tuning parameter, $\gamma$, in the form of an exponent on the mixing probability $\pi_r$.  They consider using only the values of $\gamma \in \left\{0,1/2,1\right\}$.  They suggest using the value of 0 when the true mixing proportions are not very different from each other and using 1/2 or 1 when the mixing proportions are unbalanced.  Our method corresponds to the case where $\gamma = 1$.  In other simulations (not presented here) we compared models that resulted from using different values of $\gamma$ in both balanced and unbalanced settings but generally saw little difference with respect to predictive loss when using different values for $\gamma$.

The ability to conduct inference on the estimated component coefficient functions is of critical importance.  Future work will focus on developing methods for constructing confidence bands for the the true component coefficient functions.  Subsequent work will also focus on extending the model presented here to incorporate additional scalar covariates and implementing fitting algorithms that allow for 2- and 3-dimensional images as predictors.

\section*{Acknowledgements}
This work was partially supported by NIBIB grant 5 R01 EB009744.

\bibliography{wbfmrrefs}

\end{document}